\documentclass[aip,reprint,english,a4paper]{revtex4-1}

\usepackage{upgreek}
\usepackage{amsmath}
\usepackage{amsfonts}
\usepackage{amstext}
\usepackage{mathtools}
\usepackage{amssymb}
\usepackage{siunitx}
\usepackage[usenames]{color}
\usepackage[normalem]{ulem}
\newcommand{\dirfig}{./}

\newcommand*{\diff}[1]{\mathop{\mathrm d #1}}
\newcommand*{\mean}[1]{\left< #1 \right>}
\newcommand*{\up}{\textsuperscript}

\newcommand*{\kT}{k_{\rm B}T}

\makeatletter
\let\@fnsymbol\@fnsymbol@latex
\@booleanfalse\altaffilletter@sw
\makeatother

\newcommand{\SIFEconvergence}       {1}
\newcommand{\SIJackknife}           {2}

\newcommand{\SIvelocity}     	    {4}
\newcommand{\SIdLEfriction}         {5}
\newcommand{\SITabTimestepNaCl}     {1}

\begin{document}

\title{Multisecond ligand dissociation dynamics from atomistic simulations} 

\author{Steffen Wolf}
\email[email:~]{steffen.wolf@physik.uni-freiburg.de;\\stock@physik.uni-freiburg.de}
\affiliation{Biomolecular Dynamics, Institute of Physics, Albert Ludwigs University, 79104 Freiburg, Germany}
\author{Benjamin Lickert}
\affiliation{Biomolecular Dynamics, Institute of Physics, Albert Ludwigs University, 79104 Freiburg, Germany}
\author{Simon Bray}
\affiliation{Biomolecular Dynamics, Institute of Physics, Albert Ludwigs University, 79104 Freiburg, Germany}
\affiliation{Present address: Bioinformatics Group, Department of Computer Science, Albert Ludwigs University, 79110 Freiburg, Germany}
\author{Gerhard Stock}
\email[email:~]{steffen.wolf@physik.uni-freiburg.de;\\stock@physik.uni-freiburg.de}
\affiliation{Biomolecular Dynamics, Institute of Physics, Albert Ludwigs University, 79104 Freiburg, Germany}

\begin{abstract}
  Coarse-graining of fully atomistic molecular dynamics simulations is
  a long-standing goal in order to allow the description of processes
  occurring on biologically relevant timescales. For example, the
  prediction of pathways, rates and rate-limiting steps in
  protein-ligand unbinding is crucial for modern drug discovery. To
  achieve the enhanced sampling, we first perform
  dissipation-corrected targeted molecular dynamics simulations, which
  yield free energy and friction profiles of the molecular process
  under consideration. In a second step, we use these fields to
  perform temperature-boosted Langevin simulations which account for
  the desired molecular kinetics occurring on multisecond timescales
  and beyond. Adopting the dissociation of solvated sodium chloride as
  well as trypsin-benzamidine and Hsp90-inhibitor protein-ligand
  complexes as test problems, we are able to reproduce rates from
  molecular dynamics simulation and experiments within a factor of
  2--20, and dissociation constants within a factor of 1--4. Analysis
  of the friction profiles reveals that binding and unbinding dynamics
  are mediated by changes of the surrounding hydration shells in all
  investigated systems.
\end{abstract}

\date{\today}
\maketitle

Classical molecular dynamics (MD) simulations in principle allow us to
describe biomolecular processes in atomistic detail\cite{Berendsen07}.
Prime examples include the study of protein complex
formation\cite{Pan19} and protein-ligand binding and
unbinding\cite{Bruce18,Rico19}, which constitute key
steps in biomolecular function. Apart from structural analysis, the
prediction of kinetic properties has recently become of interest,
since optimized ligand binding and unbinding kinetics have been linked
to an improved drug
efficacy\cite{Copeland06,Swinney12,Pan13,Klebe14,Copeland16}. Since
these processes typically occur on timescales from milliseconds to
hours, however, they are out of reach for unbiased all-atom MD
simulations which currently reach microsecond timescales.  To account
for rare biomolecular processes, a number of enhanced sampling
techniques\cite{Chipot07,Christ10,Mitsutake01,Torrie77,
  Isralewitz01,Sprik98,Grubmueller95,Barducci11,Comer15} have been
proposed.  These approaches all entail the application of a bias to
the system in order to enforce motion along a usually one-dimensional
reaction coordinate $x$, such as the protein-ligand distance.

While the majority of the above methods focuses on the calculation of
the stationary free energy profile $\Delta G(x)$, several approaches
have recently been suggested that combine enhanced sampling with a
reconstruction of the dynamics of the
process\cite{Tiwary13,Wu16,Teo16}. In this vein, we recently
proposed dissipation-corrected targeted MD (dcTMD), which exerts a
pulling force on the system along reaction coordinate $x$ via a moving
distance constraint\cite{Wolf18}. By combining a Langevin equation
analysis with a cumulant expansion of Jarzynski's
equality\cite{Jarzynski97}, dcTMD yields both $\Delta G(x)$ and the
friction field $\Gamma(x)$. Reflecting interactions
with degrees of freedom orthogonal to those which define the free
energy, the friction accounts for the dynamical aspects of 
the considered process.
In this work, we go one step further and use $\Delta G(x)$ and
$\Gamma(x)$ to run Langevin simulations, which describe the
coarse-grained dynamics along the reaction coordinate and reveal
timescales and mechanisms of the considered process. Moreover, we
introduce the concept of ''temperature boosting'' of the Langevin
equation, which allows us to speed up the calculations by several
orders of magnitude in order to reach biologically relevant timescales.

%
%
\section*{Theory}

\textbf{Dissipation-corrected targeted molecular dynamics.} 
To set the stage, we briefly review the working equations of dcTMD
derived in Ref.\ \onlinecite{Wolf18}. TMD as developed by Schlitter et al.
\cite{Schlitter94} uses a constraint force $f_{\rm c}$
that results in a moving distance constraint $x = x_0 + v_{\rm c} t$
with a constant velocity $v_{\rm c}$.
The main assumption underlying dcTMD is that this 
nonequilibrium process can be described by a memory-free Langevin
equation\cite{Berendsen07},
\begin{equation}\label{eq:LE}
 m \ddot x(t) = - \frac{\mathrm{d}G}{\mathrm{d}x} - \Gamma(x) \dot x +
 \sqrt{2 \kT \Gamma(x)}\, \xi (t) + f_{\rm c}(t) ,
\end{equation}
which contains the Newtonian force $- dG/dx$, the friction force
$-\Gamma(x)\dot x$, as well as a stochastic force with white noise
$\xi (t)$, that is assumed to be of zero mean,
$\langle \xi \rangle = 0$, delta-correlated,
$\langle \xi (t) \xi (t') \rangle = \delta (t-t')$, and Gaussian
distributed.
Since the constraint force $f_{\rm c}$ imposes a constant velocity on 
the system ($\dot x = v_{\rm c}$), the total force $m \ddot x$
vanishes. Performing an ensemble average $\langle \ldots \rangle$ of
Eq.\ (\ref{eq:LE}) over many TMD runs, we thus obtain the
relation \cite{Wolf18}
\begin{align}\label{eq:dG2}
 \Delta G(x) = \langle W(x) \rangle - v_{\rm c} \int_{x_0}^{x}
 \Gamma(x') \,\mathrm{d}x' .
\end{align}
Here the first term
$\langle W(x) \rangle = \int_{x_0}^{x} \langle f_{\rm c}(x') \rangle
\,\mathrm{d}x' $ represents the averaged external work performed on
the system, and the second term corresponds to the dissipated work
$W_{\rm diss}(x)$ of the process expressed in terms of the friction
$\Gamma(x)$.

While the friction in principle can be calculated in various
ways\cite{Straub87, Hummer05}, it proves advantageous to invoke
  Jarzynski's identity\cite{Jarzynski97},
  $e^{- \Delta G(x)/\kT}= \langle e^{- W(x)/\kT}\rangle$, which allows
  us to calculate $\Gamma(x)$ directly from TMD simulations. To
  circumvent convergence problems associated with the above
  exponential average \cite{Vaikuntanathan08}, we perform a
  second-order cumulant expansion which gives Eq.\ (\ref{eq:dG2})
with $W_{\rm diss}(x) = \mean{\delta W^2(x)}/\kT$. Expressing work
fluctuations $\delta W$ in terms of the fluctuating force
$\delta f_{\rm c}$, we obtain for the friction \cite{Wolf18}
\begin{align} \label{eq:dcTMD}
\Gamma (x) = \frac{1}{\kT} \int_{t_0}^{t(x)} \mean{\delta f_c(t)
  \delta f_c(t')} \diff{t'} , 
\end{align}
which is readily evaluated directly from the TMD simulations.

As discussed in Ref.\ \onlinecite{Wolf18}, the derivation of Langevin
equation (\ref{eq:LE}) assumes that the pulling speed $v_{\rm c}$ is
slow compared to the timescale of the bath fluctuations, such that the
effect of $f_{\rm c}$ can be considered as a slow adiabatic
change\cite{Servantie03}.
This means that the free energy
(\ref{eq:dG2}) and the friction (\ref{eq:dcTMD}) determined by the
nonequilibrium TMD simulations correspond to their equilibrium
results.  As a consequence, we can use $\Delta G(x)$ and $\Gamma(x)$
to describe the unbiased motion of the system via Langevin equation
(\ref{eq:LE}) for $f_{\rm c}=0$. Numerical propagation of the unbiased
Langevin equation then accounts for the coarse grained dynamics of the
system.
In this way, calculations of $\Delta G(x)$ and $\Gamma(x)$ as
well as dynamical calculations are based on the same theoretical
footing (i.e., the Langevin equation), and are therefore expected to
yield a consistent estimation of the timescales of the considered
process. Moreover, the exact solution of the Langevin equation allows
us to directly use the computed fields $\Delta G(x)$ and $\Gamma(x)$
and thus to avoid further approximations\cite{Haenggi90}.

The theory developed above rests on two main assumptions. For one, we
have assumed that the Langevin equation (\ref{eq:LE}) provides an
appropriate description of nonequilibrium TMD simulations, and applies
as well to the unbiased motion ($f_{\rm c}=0$) of the system. This
means that, due to a timescale separation of slow pulling speed and
fast bath fluctuations, the constraint force $f_{\rm c}$ enters this
equation merely as an additive term. Secondly, to ensure rapid
  convergence of the Jarzynski equation, we have invoked a cumulant
  expansion to derive the friction coefficient in Eq.\
(\ref{eq:dcTMD}), which is valid under the assumption that the
distribution of the work is Gaussian within the ensemble. While this
assumption may break down if the system of interest follows multiple
reaction paths, we have recently shown that we can systematically
perform a separation of dcTMD trajectories according to pathways by a
nonequilibrium principal component analysis of protein-ligand
contacts\cite{Post19}.  This approach bears similarities with the
  work of Tiwary et al.\ for the construction of path collective
  variables\cite{Tiwary15}.  Alternatively, path separation can be
based on geometric distances between individual trajectories, making
use of the NeighborNet algorithm\cite{Bryant04}. Details on the
  convergence of the free energy and friction estimates, the path
  separation, and the choice of the pulling velocity are given in
  Supplementary Methods and in Supplementary Figs.\ \SIFEconvergence\
  to \SIvelocity.

%
%
\begin{figure*}[htb!]
\centering
\includegraphics[width=11.4cm]{\dirfig/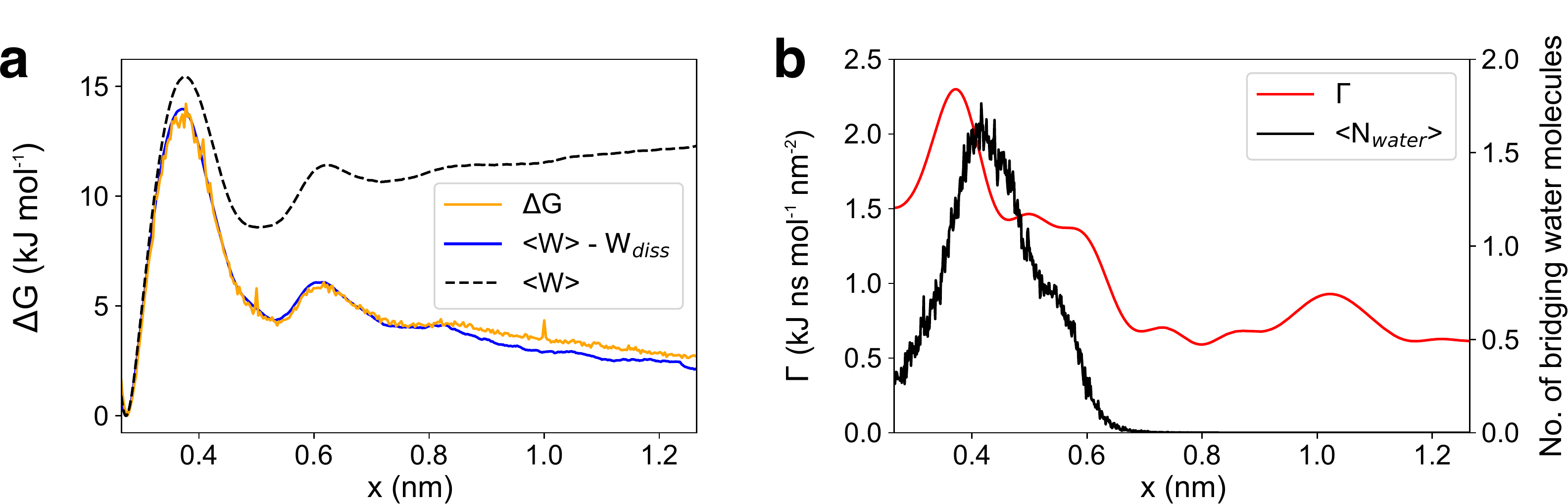}
\caption{Dissociation of NaCl in water. (\textbf{a}) Free energy
  profiles $\Delta G(x)$ along the interionic distance $x$, obtained
  from a $1\,\upmu$s long unbiased MD trajectory at 293~K (orange
  line) and $1000 \times 1\,$ns TMD runs (blue line). Error bars
    are given in Supplementary Fig.\ \SIJackknife. Also shown is the
  average work $\langle W(x) \rangle$ calculated from the TMD
  simulations (dashed black line). (\textbf{b}) Friction profile
  $\Gamma(x)$ (red) obtained from dcTMD after Gaussian smoothing
  together with the average number of water molecules (black),
  that connect the Na$^+$ and Cl$^-$ ions in a common hydration
  shell \cite{Mullen14}.}
\label{fig:NaCl}
\end{figure*}

\textbf{$T$-boosting.}
The speed-up of Langevin equation (\ref{eq:LE}) compared to an
unbiased all-atom MD simulation is due to the drastic coarse graining
of the Langevin model (one instead of $3N$ degrees of freedom, $N$
being the number of all atoms). Since the numerical integration of the
Langevin equation typically requires a time step of a few femtoseconds (see
Supplementary Table \SITabTimestepNaCl), however, we still need to propagate Eq.\
(\ref{eq:LE}) for $ \gtrsim 100 \cdot 10^{15}$ steps to
sufficiently sample a process occurring on a timescale of
seconds, which is prohibitive for standard computing resources.

As a further way to speed up calculations, we note that the
temperature $T$ enters Eq.\ (\ref{eq:LE}) via the stochastic force,
indicating that temperature is the driving force of the Langevin
dynamics. That is, when we consider a process described by a
transition rate $k$ and increase the temperature from $T_1$ to $T_2$,
the corresponding rates $k_1$ and $k_2$ are related by the
Kramers-type expression\cite{Haenggi90}
\begin{equation} \label{eq:rates}
k_2 = k_1 e^{- \Delta G^{\neq}(\beta_2 -\beta_1)} ,
\end{equation}
where $\Delta G^{\neq}$ denotes the transition state energy and
$\beta_i = 1/k_{\rm B}T_i$ is the inverse temperature. Hence, by
increasing the temperature we also increase the number $n$ of observed
transition events according to $n_2/n_1 = k_2/k_1$.
%

To exploit this relationship for dcTMD, we proceed as follows.  First
we employ dcTMD to calculate the Langevin fields $\Delta G(x)$ and
$\Gamma(x)$ at a temperature of interest $T_1$. Using these fields, we
then run a Langevin simulation at some higher temperature $T_2$, which
results in an increased transition rate $k_2$ and number of events
$n_2$.  In particular, we choose a temperature high enough to sample a
sufficient number of events ($N \gtrsim 100$) for some given
simulation length. In the final step, we use Eq.\ (\ref{eq:rates}) to
calculate the transition rate $k_1$ at the desired temperature $T_1$.

As Eq.\ (\ref{eq:rates}) arises as a consequence\cite{Haenggi90} of
Langevin equation (\ref{eq:LE}), the above described procedure,
henceforth termed ``$T$-boosting,'' involves no further
approximations. It exploits the fact that we calculate fields
$\Delta G(x)$ and $\Gamma(x)$ at the same temperature for which we
eventually want to calculate the rate.
We wish to stress that this virtue represents a crucial
  difference to temperature accelerated MD.\cite{Sorensen00} In the
  latter method the
  free energy $\Delta G(x)$ is first calculated at a high temperature
  and subsequently rescaled to a desired low temperature, whereupon
  $\Delta G(x)$ in general does change. $T$-boosting avoids this,
  because by using dcTMD we calculate $\Delta G(x)$ right away at the
  desired temperature.
We note in passing that
a Langevin simulation run at $T_2$ using fields obtained at $T_1$ in
general does not reflect the coarse-grained dynamics of an MD
simulation run at $T_2$, but can only be used to recover
$k_1$ from $k_2$.

In practice, we perform $T$-boosting calculations at several
temperatures $T_2$ in increments of 25~K to 50~K and choose the
smallest $T_2$ such that $N \gtrsim 100$ transitions occur.  In the
Supporting Information we derive an analytic expression of the
extrapolation error as a function of boosting temperatures and
achieved number of transitions, from which the necessary length of the
individual Langevin simulations can be estimated, in order to achieve
a desired extrapolation error. One-dimensional Langevin simulations
require little computational effort (1~ms of simulation time at a 5~fs
time step take $\sim$6 hours of wall-clock time on a single CPU) and
are trivial to parallelize in the form of independent short runs. Hence
the extrapolation error due to boosting can easily be pushed below
10\% and is thus negligible in comparison to systematic errors coming from
the dcTMD field estimates. 

  As shown in Supplementary Table \SITabTimestepNaCl, a further
  increase in efficiency can be achieved if the considered dynamics is
  overdamped, which is the case for both protein-ligand systems.
  Since overdamped dynamics neglects the inertia term $m \ddot x$ and
  therefore does not depend on the mass $m$, we may artificially
  enhance the mass in the Langevin simulations. For the protein-ligand
  systems, this allows us to increase the integration time step
  from 1 to 10 fs, i.e., a speed-up of an order of
  magnitude.

%
%
\section*{Results and Discussion}

\begin{figure*}[htb!]
\centering
\includegraphics[width=16.5cm]{\dirfig/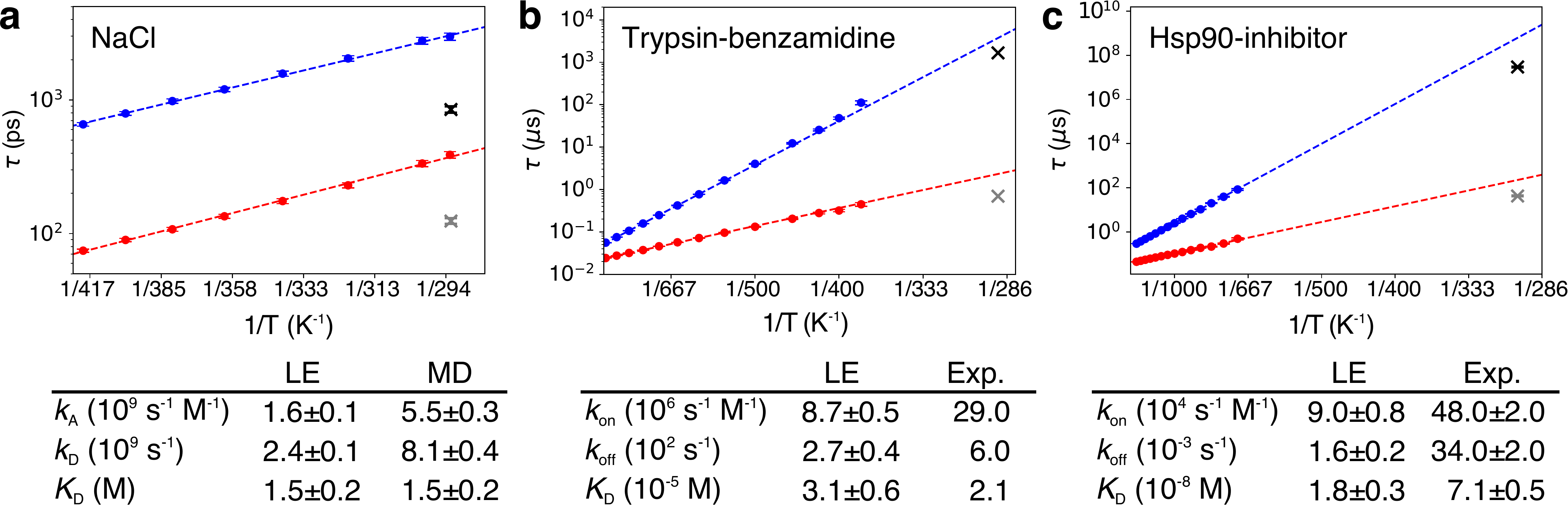} 
\caption{Mean binding (red) and unbinding (blue) times, drawn as a
  function of the inverse temperature, obtained from $T$-boosted
  Langevin simulations of (\textbf{a}) solvated NaCl, (\textbf{b}) the
  trypsin-benzamidine complex and (\textbf{c}) the Hsp90-inhibitor
  complex. Dashed lines represent fits (R$^2 = 0.90-0.99$) to Eq.\
  (\ref{eq:rates}), crosses (binding in grey, unbinding in black)
  indicate reference results from \textbf{a} unbiased MD
  simulation\cite{Wolf18} and \textbf{b}, \textbf{c}
  experiment\cite{Guillain70,Amaral17}. Bars represent the standard
  error of the mean. Tables below comprise corresponding rates (with
  M being the molarity, i.e., mol/l) and reference values.  Rate
  constants were fitted according to Eq.\ (\ref{eq:rates}) at $290$~K
  and $300$~K, respectively, with fit errors as
  indicated\cite{Hughes10}. Dissociation constants were calculated
  from rate constants.}
\label{fig:accelerate}
\end{figure*}

\textbf{Ion dissociation of NaCl in water.}
To illustrate the above developed theoretical concepts and test the
validity of the underlying approximations, we first consider sodium
chloride in water as a simple yet nontrivial model system. For this system,
detailed dcTMD as well as long unbiased MD simulations are
available\cite{Wolf18}, making it a suitable benchmark system for our approach. 

Fig.\ \ref{fig:NaCl}a shows the
free energy profiles $\Delta G(x)$ along the interionic distance $x$,
whose first maximum at $x \approx 0.4$~nm corresponds to the
binding-unbinding transition of the two ions. 
The second smaller maximum at $x \approx 0.6$~nm reflects the
transition from a common to two separate hydration
shells\cite{Mullen14}. We find that results for $\Delta G(x)$ obtained
from a $1\,\upmu$s long unbiased MD trajectory and from dcTMD
simulations ($1000 \times 1\,$ns runs with $v_{\rm c}\!=\!1\,$m/s)
match perfectly. Since the average work $\langle W(x) \rangle$ of the
nonequilibrium simulations is seen to significantly overestimate the
free energy at large distances, the dissipation correction
$W_{\rm diss}$ in Eq.\ (\ref{eq:dG2}) is obviously of importance.
Figure \ref{fig:NaCl}b shows the underlying friction profile
$\Gamma(x)$ obtained from dcTMD, which in part deviates from the
lineshape of the free energy. While we also find a maximum at
$x \approx 0.4$~nm, the behavior of $\Gamma(x)$ is remarkably
different for larger distances $0.5 \lesssim x \lesssim 0.7$~nm, where
a region of elevated friction can be found before dropping to lower
values. Interestingly, these features of
$\Gamma(x)$ match well the changes of the average number of water
molecules bridging both ions\cite{Mullen14}. This indicates that the increased
friction in Eq.\ (\ref{eq:dcTMD}) is mainly caused by force
fluctuations associated with the build-up of a hydration
shell\cite{Wolf18}. For $x \gtrsim 0.8$~nm, the friction is constant within our
signal-to-noise resolution.

The dynamics of ion dissociation and association can be described by
their mean waiting times and corresponding rates shown in
Figure~\ref{fig:accelerate}a. For the chosen force field, ion
concentration and resulting effective simulation box size,
the unbiased MD simulation at 293 K yields mean dissociation and
association times of $\tau_{\rm D}=1/k_{\rm D}=120\,$ps and
$\tau_{\rm A}=1/\left( k_{\rm A} C \right)=850\,$ps, respectively,
where $C$ denotes a reference
concentration (see the Supplementary Methods for details). Using
fields $\Delta G(x)$ and $\Gamma(x)$ obtained from TMD, the numerical
integration of Langevin equation (\ref{eq:LE}) for 1 $\upmu$s results
in $\tau_{\rm D}=420\,$ps and $\tau_{\rm A}=3040\,$ps. While the
dissociation constants $K_{\rm D} = k_{\rm D} / k_{\rm A} = 1.5$~M from
Langevin and MD simulations match perfectly, we find that the Langevin
predictions overestimate the correct rates by a factor of
$\sim$3.4. The latter may be caused by various issues.
For one, to be of practical use, the Langevin model was deliberately
kept quite simple. For example, it does not include an explicit
solvent coordinate\cite{geissler99,Mullen14}, but accounts for the
complex dynamics of the solvent merely through the friction field
$\Gamma(x)$. Moreover, we note that the calculation of $\Gamma(x)$
via Eq.\ (\ref{eq:dcTMD}) uses constraints, which have the effect of
increasing the effective friction\cite{Daldrop17}. This finding is
supported by calculations using the data-driven Langevin
approach\cite{Hegger09,Schaudinnus16}, which estimates friction coefficients
based on unbiased MD simulations that are consistantly smaller than the ones
obtained from dcTMD (Supplementary
Fig.~\SIdLEfriction).
Considering the simplicity of the Langevin model and the approximate
calculation of the friction coefficient by dcTMD, overall we are content
with a factor $\sim 3$ deviation of the predicted kinetics. 

To illustrate the validity of the
$T$-boosting approach suggested above, we performed a series
of Langevin simulations for eight temperatures ranging from 290 to
420~K and plotted the resulting dissociation and association times as
a function of the inverse temperature (Fig.\ \ref{fig:accelerate}a). 
Checking the consistency of our approach, a
fit to Eq.\ (\ref{eq:rates}) yields transition state free energies
$\Delta G^{\neq}$ of $13$~kJ/mol and $12$~kJ/mol for ion dissociation
and association, respectively, which agree well with barrier heights
of the free energy profile in Fig.\ \ref{fig:NaCl}a.
Moreover, dissociation and association times obtained from the
extrapolated $T$-boosted Langevin simulations ($\tau_{\rm D}=370$~ps,
$\tau_{\rm A}=3050$~ps) agree excellently with the directly calculated
values. This
indicates that high-temperature Langevin simulations can indeed be
extrapolated to obtain low-temperature transition rates.

%
%
  
\textbf{Trypsin-benzamidine.}
Let us now consider the prediction of
free energies, friction profiles and kinetics in protein-ligand
systems.  The first system we focus on is the inhibitor benzamidine
bound to trypsin\cite{Guillain70,Marquart83,Schiebel18}, which
represents a well-established model problem to test enhanced sampling
techniques\cite{Buch11,Plattner15,Tiwary15,
  Teo16,Votapka17,Betz19}. The slowest dynamics in this system is
found in the unbinding process, which occurs on a scale of
milliseconds\cite{Guillain70}. To capture the kinetics of the
unbinding process, so far Markov state models\cite{Buch11,Plattner15},
metadynamics\cite{Tiwary15}, Brownian dynamics\cite{Votapka17} and
adaptive enhanced sampling methods\cite{Teo16,Betz19} have been
employed.

\begin{figure*}[htb]
\centering
\includegraphics[width=15.8cm]{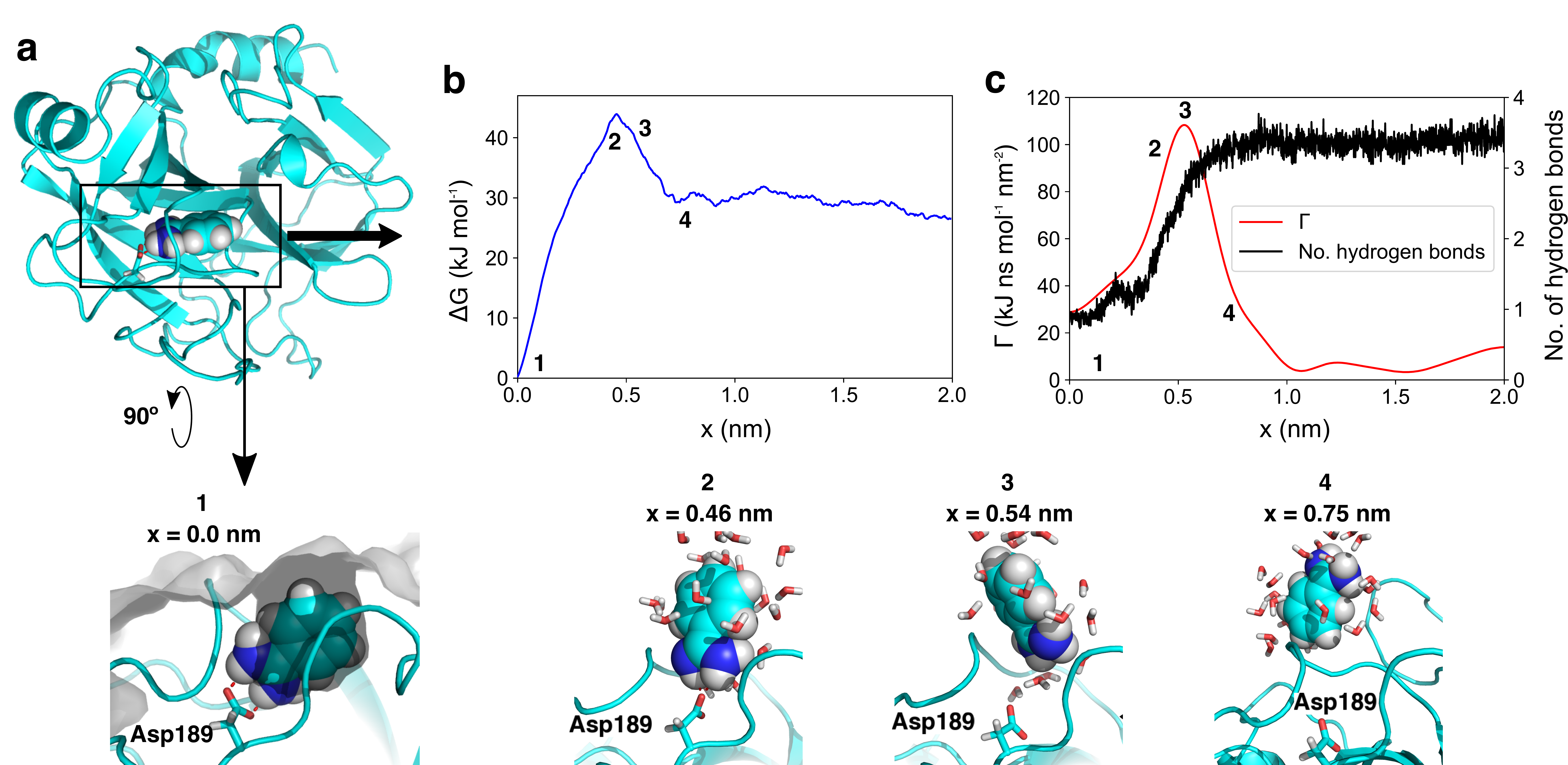} 
\caption{Unbinding of benzamidine from trypsin. (\textbf{a}) TMD
  snapshots of the structural evolution in trypsin along the dominant
  dissociation pathway, showing protein surface in gray, benzamidine
  as van der Waals spheres, Asp189 and water molecules as
  sticks. Benzamidine is bound to the protein in a cleft of the
  protein surface via a bidental salt bridge to Asp189. dcTMD
  calculations of (\textbf{b}) free energy $\Delta G(x)$ and (\textbf{c})
  (Gaussian smoothed) friction $\Gamma(x)$ together with the mean
  number of hydrogen bonds between benzamidine and water. Highlighted
  are the bound state \textbf{1}, transition state \textbf{2}, the
  state with maximal friction \textbf{3} and the unbound state
  \textbf{4}. Error bars of free energy and friction estimates
    are given in Supplementary Fig.\ \SIJackknife.}
   \label{fig:Tryps}
\end{figure*}

Here we combined dcTMD simulations and a subsequent nonequilibrium
principal component analysis\cite{Post19} to identify the dominant
dissociation pathways of ligands during unbinding from their host
proteins (see Supplementary Methods). Figure~\ref{fig:Tryps} shows TMD
snapshots of the structural evolution along this pathway, its free
energy profile $\Delta G(x)$, and the associated friction
$\Gamma(x)$. Starting from the bound state ($x_1\!=\!0$~nm), $\Delta G(x)$
exhibits a single maximum at $x_2\approx 0.46$~nm, before it reaches
the dissociated state for $x\gtrsim x_4= 0.75$~nm.
%
%
In line
with the findings of Tiwary et al.\cite{Tiwary15}, the maximum of
$\Delta G(x)$ reflects the rupture of the Asp189-benzamidine salt
bridge, which represents the most important contact of the bound
ligand. Following right after, the friction profile $\Gamma(x)$
reaches its maximum at $x_3 \approx 0.54$~nm, where the charged side
chain of benzamidine becomes hydrated with water molecules. Similarly
to NaCl, the friction peak coincides with the increase in the average
number of hydrogen bonds between benzamidine and bulk water.  The peak
in friction is slightly shifted to higher $x$, because the ligand acts
as a ''plug'' for the binding site, and first needs to be (at least
partially) removed in order to allow water flowing in. As for the
dissociation of NaCl in water, enhanced friction during unbinding
appears to be directly linked to a rearrangement of the protein-ligand
hydration shell, which is in agreement with recent results from
neutron crystallography\cite{Schiebel18}.
 
To calculate rates $k_{\rm on}$ and $k_{\rm off}$ describing the
binding and unbinding of benzamidine from trypsin, we performed 10~ms
long Langevin simulations along the dominant pathways at thirteen
temperatures ranging from 380--900 K. As shown in
Fig.~\ref{fig:accelerate}b, the resulting rates are well fitted
(R\up{2} $\geq 0.90$) by the $T$-boosting expression in Eq.\
(\ref{eq:rates}). Representing the resulting number of transitions as
a function of the inverse temperature, we find that at 380 K only
$\sim 9$ events happen during a millisecond. That is, to obtain
statistically converged rates at 290 K would require Langevin
simulations at 290 K on a timescale of seconds. Using temperature
boosting with Eq.\ (\ref{eq:rates}), on the other hand, our
high-temperature millisecond Langevin simulations readily yield
converged transition rates at 290 K (see Fig.\ \ref{fig:accelerate}b),
that is, $k_{\rm on}= 8.7 \cdot 10^{6}$~s\up{-1}M\up{-1} and
$k_{\rm off}= 2.7 \cdot 10^{2}$~s\up{-1}, which underestimate the
experimental values\cite{Guillain70}
$k_{\rm on}= 2.9 \cdot 10^{7}$~s\up{-1}M\up{-1} and
$k_{\rm off}= 6.0 \cdot 10^{2}$~s\up{-1} by a factor of 2--3.
Similarly, the calculated $K_{\rm D}$ overestimates the
  experimental result \cite{Guillain70} of
  $K_{\rm D} = 2.1 \cdot 10^{-5}$~M by a factor of $\sim$1.5. As
  indicated by a recent review\cite{Bruce18} comparing numerous
  computational methods to calculate (un)binding rates of
  trypsin-benzamidine, our approach compares quite favorably regarding
  accuracy and computational effort.

As the extrapolation error due to $T$-boosting is negligible (see
Supplementary Methods), the observed error is mainly caused by the
approximate calculation of free energy and friction fields by dcTMD.
In the case of NaCl, we have shown that reliable estimates of the
fields (with errors $\lesssim 1\, \kT$) require an ensemble of at
least 500 simulations (see Ref.~\onlinecite{Wolf18} and
Supplementary Fig.\ \SIJackknife), although the means of $\Delta G$
and $\Gamma$ appear to converge already for $\sim$100 trajectories. In
a similar vein, by performing a Jackknife ''leave-one-out''
analysis\cite{Efron81}, for trypsin-benzamidine we obtain an error of
$\sim 2\, \kT$ for 150 trajectories (Supplementary Fig.\
\SIJackknife). Interestingly, the error of the main free energy
barrier is typically comparatively small, because the friction and
thus variance of $W$ increase directly after the barrier. As a
consequence, the sampling error of $k_{\rm off}$ is small compared to
that of $k_{\rm on}$ and the binding free energy. We note that if the
experimental binding affinity $K_D$ is known, it can be used as a
further constraint on the error of the free energy and friction
fields.

\begin{figure*}[ht]
\centering
\includegraphics[width=15.8cm]{\dirfig/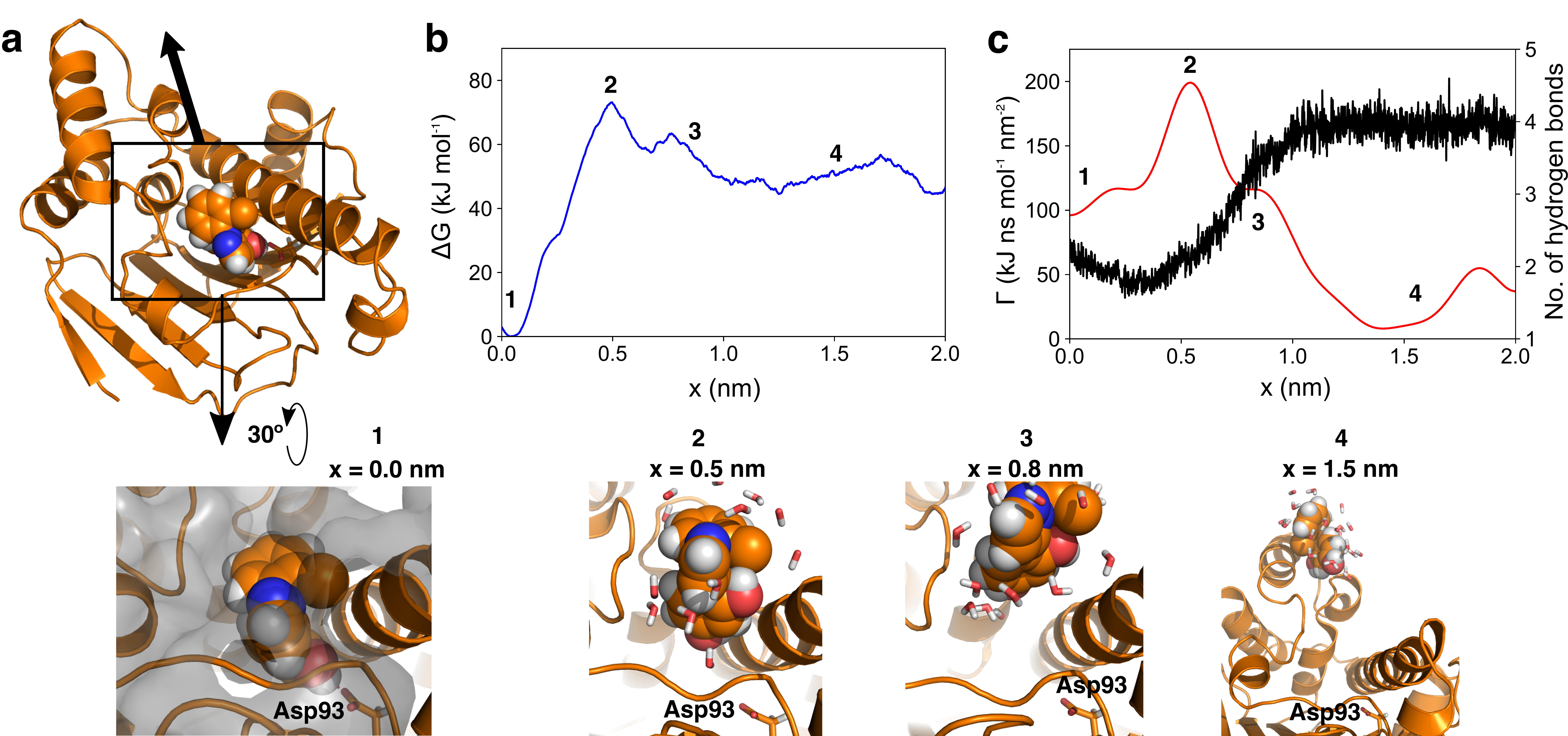} 
\caption{Unbinding of an inhibitor from the N-terminal domain of
  Hsp90. (\textbf{a}) Structural evolution along the dissociation
  pathway in Hsp90, showing protein surface in gray, inhibitor as van
  der Waals spheres, Asp93 and water molecules as sticks. The
  inhibitor is bound to the protein in a cleft of the protein surface
  via a hydrogen bond to Asp93. dcTMD calculations of (\textbf{b}) free
  energy $\Delta G(x)$ and (\textbf{c}) (Gaussian-smoothed) friction
  $\Gamma(x)$ together with the mean number of hydrogen bonds
  between inhibitor and water. Highlighted are the bound state
  \textbf{1}, transition state and state with maximal friction
  \textbf{2}, an additional state with increased friction \textbf{3}
  and the unbound state \textbf{4}. Error bars of free energy and
    friction estimates are given in Supplementary Fig.\ \SIJackknife.
  Fluctuations of $\Gamma(x)$ for $x \gtrsim 1\,$nm are due to
  noise. Color code as in Fig.~\ref{fig:Tryps}.}
   \label{fig:Hsp90}
\end{figure*}

%
%
\textbf{Hsp90-inhibitor.}
The second investigated protein complex is
the N-terminal domain of heat shock protein 90 (Hsp90) bound to a
resorcinol scaffold-based inhibitor (\textbf{1j} in
Ref.~\onlinecite{Wolf19c}). This protein has recently been established
as a test system for investigating the molecular effects influencing
binding kinetics\cite{Amaral17,Kokh18,Schuetz18,Wolf19c}, and the
selected inhibitor unbinds on a scale of half a minute. From the
overall appearance of free energy and friction profiles
(Fig.~\ref{fig:Hsp90}), we observe clear similarities to the case of
trypsin-benzamidine. That is, the main transition barrier is also
found at $x_2\approx 0.5$~nm, which stems from the ligand pushing
between two helices at this point in order to escape the binding site.
Moreover, the friction peaks at $x_2\approx 0.5$~nm, as well, but with
an additional shoulder at $x_3\approx 0.8$~nm, which again coincides
with changes of the ligand's hydration shell. The unbound state is
reached after $x \gtrsim 1.0$~nm.  We note that the ligand is again
bound to the protein via a hydrogen bond to an aspartate (Asp93) and
at a position that is open to the bulk water. 

To calculate rates $k_{\rm on}$ and $k_{\rm off}$, we again performed
5~ms long Langevin simulations along the dissociation pathway at
fourteen different temperatures ranging from 700--1350~K.  Rate
prediction yields $k_{\rm on}\!=\! 9.0\cdot 10^{4}$~s\up{-1}M\up{-1} and
$k_{\rm off}\!=\! 1.6\cdot 10^{-3}$~s\up{-1}, and underestimates the
experimental\cite{Amaral17} values
$k_{\rm on}= 4.8 \pm 0.2 \cdot 10^{5}$~s\up{-1}M\up{-1} and
$k_{\rm off}= 3.4 \pm 0.2 \cdot 10^{-2}$~s\up{-1} by a factor of
5--20.
The resulting value for $K_{\rm D} = 1.8 \cdot 10^{-8}$~M
  underestimates the experimental value\cite{Amaral17} $7.1 \cdot
  10^{-8}$~M by a factor of $\sim$4. 
Considering that we attempt to
predict unbinding times on a time scale of half a minute from
sub-$\upmu$s MD simulations, and that a factor 20 corresponds to a 
free energy difference of about 3 $k_B T$ (i.e., 15 \% of the
barrier height in Hsp90), we find this agreement remarkable for a
first principles approach which
implies many uncertainties of the physical model\cite{Capelli20}.
We attribute the larger deviation in
comparison to trypsin to issues with the sampling of the correct
unbinding pathways: especially unbinding rates in the range of minutes
to hours fall into the same timescale as slow conformational dynamics
of host proteins\cite{Amaral17}, requiring a sufficient sampling of
the conformational space of the protein as a prerequisite for dcTMD
pulling simulations.
 
%
%
%
\section*{Conclusions}

Using free energy and friction profiles obtained from dcTMD, we have
shown that $T$-boosted Langevin simulations yield binding and
unbinding rates which are well comparable to results from atomistic
equilibrium MD and experiments.
That is, rates are underestimated by
an order of magnitude or less which, in comparison to other methods
that have been applied to the trypsin-benzamidne and Hsp90 complexes
(see Refs.\ \onlinecite{Bruce18,NunesAlves20} for recent reviews), is
within the top accuracy currently achievable. At the same time, the
few other methods that aim at predicting absolute rates (such as
Markov state models\cite{Buch11,Plattner15} and infrequent
metadynamics\cite{Tiwary15,Casasnovas17}) require substantial more MD
simulation time, while dcTMD only requires
sub-$\upmu$s MD runs, that is, at least an order of magnitude less
computational time.
As the extrapolation error due to $T$-boosting is
negligible, the error is mainly caused by the approximate calculation
of free energy and friction fields by dcTMD. We have shown that
friction profiles, which correspond to the dynamical aspect of
ligand binding and unbinding, may yield additional insight into
molecular mechanisms of unbinding processes, which are not reflected
in the free energies. Although the three investigated molecular
systems differ significantly, in all cases friction was found to be
governed by the dynamics of solvation shells.

%
%
\section*{Methods}

\textbf{MD simulations.}
All simulations employed Gromacs v2018 (Ref. \onlinecite{Abraham15})
in a CPU/GPU hybrid implementation, using the Amber99SB* force
field\cite{Hornak06,Best09} and the TIP3P water
model\cite{Jorgensen83}. For each system, $10^2$-$10^3$ dcTMD
calculations\cite{Wolf18} at pulling velocity $v_c = 1$~m/s were
performed to calculate free energy $\Delta G(x)$ and friction
$\Gamma (x)$. For the NaCl-water system, dcTMD as well as unbiased MD
simulations were taken from
Ref.~\onlinecite{Wolf18}. Trypsin-benzamidin complex simulations are
based on the 1.7~\AA~X-ray crystal structure with PDB ID 3PTB
(Ref. \onlinecite{Marquart83}). Simulation systems of the
Hsp90-inhibitor complex were taken from
Ref.~\onlinecite{Wolf19c}. Detailed information on system preparation,
ligand parameterization, MD simulations and pathway separation can be
found in the Supplementary Information.

\textbf{Langevin simulations.}
Langevin simulations employed the
integration scheme by Bussi and Parrinello\cite{Bussi07b}.
Details on the performance of this method with respect to the
employed integration time step and system mass can be found in the
Supplementary Information.

\subsection*{Data availability}
\vspace*{-4mm} Python scripts for dcTMD calculations, the fastpca
program package for nonequilibrium principal component analysis, the
data-driven Langevin package, the Langevin simulation code, and
Jupyter notebooks for $T$-boosting analysis and sampling error
estimation in Langevin simulations are available at our website
\url{www.moldyn.uni-freiburg.de}. Further data is available from the
authors upon request.

%
%
\subsection*{Acknowledgements}
\vspace*{-4mm}
We thank Peter Hamm and Matthias Post for numerous instructive and helpful
discussions. The authors acknowledge support by the Deutsche
Forschungsgemeinschaft (Sto 247/11), by the bwUniCluster computing
initiative, the High Performance and Cloud Computing Group at the
Zentrum f\"ur Datenverarbeitung of the University of T\"ubingen, the
state of Baden-W\"urttemberg through bwHPC and the Deutsche
Forschungsgemeinschaft through grant No. INST 37/935-1 FUGG.

\subsection*{Author contributions}
\vspace*{-4mm}
S.W. and G.S. designed and supervised
research. S.W. performed TMD and Langevin simulations and
nonequilibrium path separation of Trypsin trajectories. B.L. performed
dLE analysis and implemented Langevin simulations. S.B. performed
the nonequilibrium path separation of Hsp90 trajectories. All authors
wrote the manuscript.

\textbf{Correspondence and requests for materials} should be addressed
to S.W. and G.S.


\newpage

\onecolumngrid
\appendix

\renewcommand{\thepage}{\arabic{chapter}.\arabic{page}}  
\renewcommand{\thesection}{\arabic{chapter}.\arabic{section}}   
\renewcommand{\thetable}{\arabic{chapter}.\arabic{table}}   
\renewcommand{\thefigure}{\arabic{chapter}.\arabic{figure}}
\renewcommand{\thepage}{S\arabic{page}}  
\renewcommand{\thesection}{S\arabic{section}}   
\renewcommand{\thetable}{S\arabic{table}}   
\renewcommand{\thefigure}{S\arabic{figure}}

\section*{Supplementary Information}

\section*{Supplementary Methods}

\baselineskip5.4mm 

\subsection*{MD simulation details.}
Protein and ion interactions were described by the Amber99SB* force
field\cite{Hornak06,Best09}, water molecules with the TIP3P
model\cite{Jorgensen83}. Simulations were carried out using Gromacs
v2018 (Ref.~\onlinecite{Abraham15}) in a CPU/GPU hybrid
implementation.  Protein protonation states were evaluated with
propka\cite{Olsson11}. Van der Waals interactions were calculated with
a cut-off of 1~nm, electrostatic interactions using the particle mesh
Ewald method\cite{Darden93} with a minimal real-space cut-off of
1~nm. All covalent bonds with hydrogen atoms were constrained using
LINCS\cite{Hess97}. After an initial steepest descent minimisation
with positional restraints on protein and ligand heavy atoms, an
initial 0.1~ns equilibration MD simulation in the NPT ensemble was
performed with a 1~fs time step and positional restraints of protein
and ligand heavy atoms. A temperature of 290.15~K was kept constant
using the Bussi (v-rescale) thermostat\cite{Bussi07a} (coupling time
constant of 0.2~ps), the pressure was kept constant at 1~bar using the
Berendsen barostat\cite{Berendsen84} (coupling time constant of
0.5~ps), followed by a second steepest descent minimisation without
restraints and a short 0.1~ns equilibration MD simulation in the NPT
ensemble.

dcTMD calculations\cite{Wolf18} were carried out using the PULL code
implemented in Gromacs using the ''constraint'' option employing a
SHAKE implementation\cite{Ryckaert77}. 
200--400 statistically independent start points of simulations
were obtained by generating different atomic velocity distributions
after the 10~ns unbiased simulations, all corresponding to a
temperature of 290.15~K. After a 0.1~ns preequilibration using
parameters as described above with positional restraints on protein
and ligand heavy atoms and a constant distance constraint of all 
simulation systems, constant velocity calculations were carried out
with $v_c = 1$~m/s covering a distance of 2~nm, switching the barostat
to the Parrinello-Rahman barostat\cite{Parrinello81}. The constraint
pseudo-force $f_c$ was written out each time step.

\textbf{NaCl.}
Free energy $\Delta G(x)$ and friction profiles $\Gamma (x)$ were
obtained from 1000 trajectories of previous dcTMD
calculations\cite{Wolf18} at pulling velocity $v_c = 1$~m/s. For a
better sampling, we continued the unbiased fully atomistic simulations
described in Ref.~\onlinecite{Wolf18} and extended them to a full
microsecond of simulated time. As these simulations used a cubic
simulation box, binding and unbinding waiting times cannot directly be
compared to the results of our Langevin simulation with ''reflective''
borders (see below), which represent radial dynamics. To obtain data
sets that allow such a comparison, we removed all time steps with
$x<0.265$~nm and $x>1.265$~nm from MD trajectories, and calculated
mean waiting times for the resulting cut $x(t)$ trajectories.

\textbf{Trypsin-benzamidine.}  
Benzamidine parameters were obtained
using Antechamber\cite{Wang06} and Acpype\cite{Sousada12} with atomic
parameters derived from GAFF parameters\cite{Wang04}. Atomic charges
were obtained as RESP charges\cite{Bayly93} based on QM calculations
at the HF/6-31G* level using Orca\cite{Neese12} and
Multiwfn\cite{Lu12}. Trypsin (PDB ID 3PTB)\cite{Marquart83} was placed
into a dodecahedral box with dimensions of 7.5 x 7.5 x 5.3~nm\up{3}
side length and solvated with 8971 water molecules. 16 sodium and 25
chloride ions were added to yield a charge neutral box with a salt
concentration of 0.1~M\cite{Guillain70}. After the initial
equilibration, we added an additional 10~ns unbiased MD simulation
to yield a converged protein structure. As pulling coordinates, we
used the distance between the center of mass of all benzamidine heavy
atoms and the one of the C$_{\alpha}$ atoms of the central
$\beta$-sheet of trypsin.

\textbf{Hsp90-inhibitor.}  
Parameters of the resorcinol inhibitor were
taken from Ref.~\onlinecite{Wolf19c}: here, inhibitor parameters were
generated using Antechamber\cite{Wang06} and Acpype\cite{Sousada12}
with atomic parameters derived from GAFF parameters\cite{Wang04} and
AM1-BCC atomic charges\cite{Jakalian00,Jakalian02}. Solvated
simulation boxes of the Hsp90-inhibitor complex were taken from
Ref.~\onlinecite{Wolf19c} (compound \textbf{1j}), which in turn are
based on the 2.5~\AA~X-ray crystal structure with PDB ID
6FCJ\cite{Guldenhaupt18}. As in the case of trypsin, the distance
between the center of mass of all ligand heavy atoms and the one of
the C$_{\alpha}$ atoms of the central $\beta$-sheet of Hsp90 served as
as pulling coordinate as used in Ref.~\onlinecite{Wolf19c}.

\textbf{Data evaluation.}
Minimal distance evaluation for contact determination was performed using the MDanalysis Python
library\cite{Michaud11}, nonequilibrium principal component analysis
was carried out using the fastpca program\cite{Sittel14}. Data
evaluation was carried out using a Jupyter notebook\cite{Kluyver16}
employing the numpy\cite{vanderWalt11}, scipy\cite{Virtanen20} and
astropy\cite{astropy18} Python libraries. Graphs were plotted using
the matplotlib\cite{Hunter07} Python library, molecular structures
were displayed via PyMOL\cite{PyMOL}.

%
%

\begin{figure*}[ht!]
\centering
\includegraphics[width=0.45\textwidth]{\dirfig/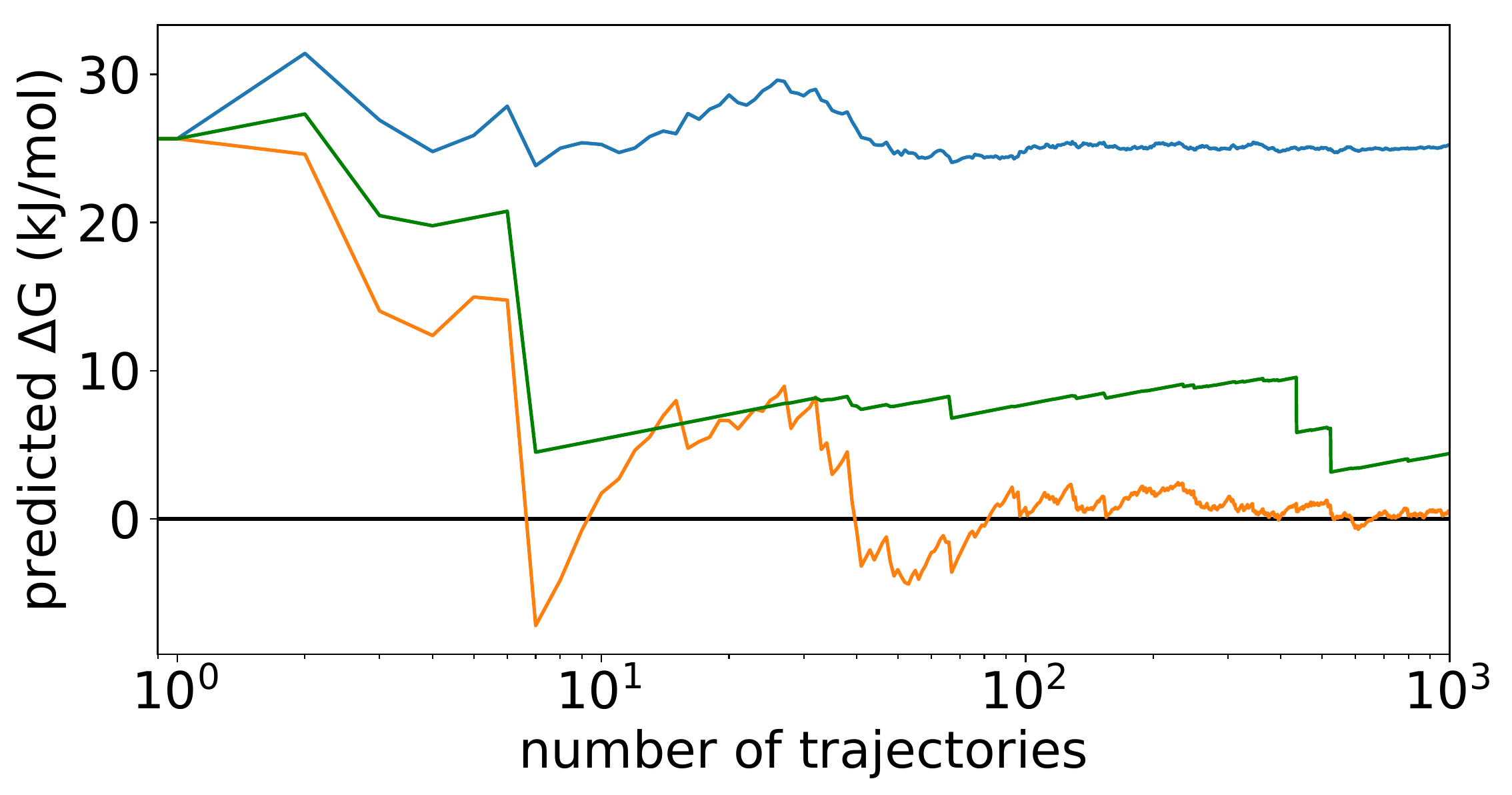} 
\caption{\baselineskip4mm \textbf{Convergence of dcTMD free energy
    estimate,} illustrated for a Gaussian work distribution.  Compared
  are $\mean{W}$ in blue, $\Delta G$ estimate by Jarzynski's identity
  in green, cumulant expansion estimate in orange. }
\label{fig:estimators}
\end{figure*}
\vspace*{-8mm}

\subsection*{Statistical convergence of dcTMD free energy and friction
  estimates} 

To illustrate the statistical convergence of various quantities (such
as mean work and free energy) calculated via the second-order cumulant
expansion of Jarzynski's identity, we have performed a detailed study
for a large range of ensemble sizes and pulling velocities in the case
of NaCl (see the Supplementary Information of Ref.\
\onlinecite{Wolf18}). Using a simple model problem, here we restrict
us to demonstrate the clear improvement of the convergence behavior
when the second-order cumulant expansion instead of the direct
evaluation of Jarzynski's identity is used.
To this end, we generated test data in form of draws from a normal
distribution with mean $\mean{W}$ and variance $\mean{\delta W^2}$
chosen such that
$\Delta G = \mean{W} -\frac{1}{2 \kT} \mean{\delta W^2} = 0$. As
Supplementary Fig.~\ref{fig:estimators} shows, the estimator from
Jarzynski's identity exhibits a slow, erratic convergence
behavior. On the other hand, the cumulant expansion-based estimator
gives the dissipated work in terms of the variance of the
nonequilibrium work, which is much easier to compute and converges to
a fluctuation around $1 \kT \approx 2.5$~kJ/mol after $\sim 100$
draws.  
\begin{figure*}[h]
\centering
\includegraphics[width=0.7\textwidth]{\dirfig/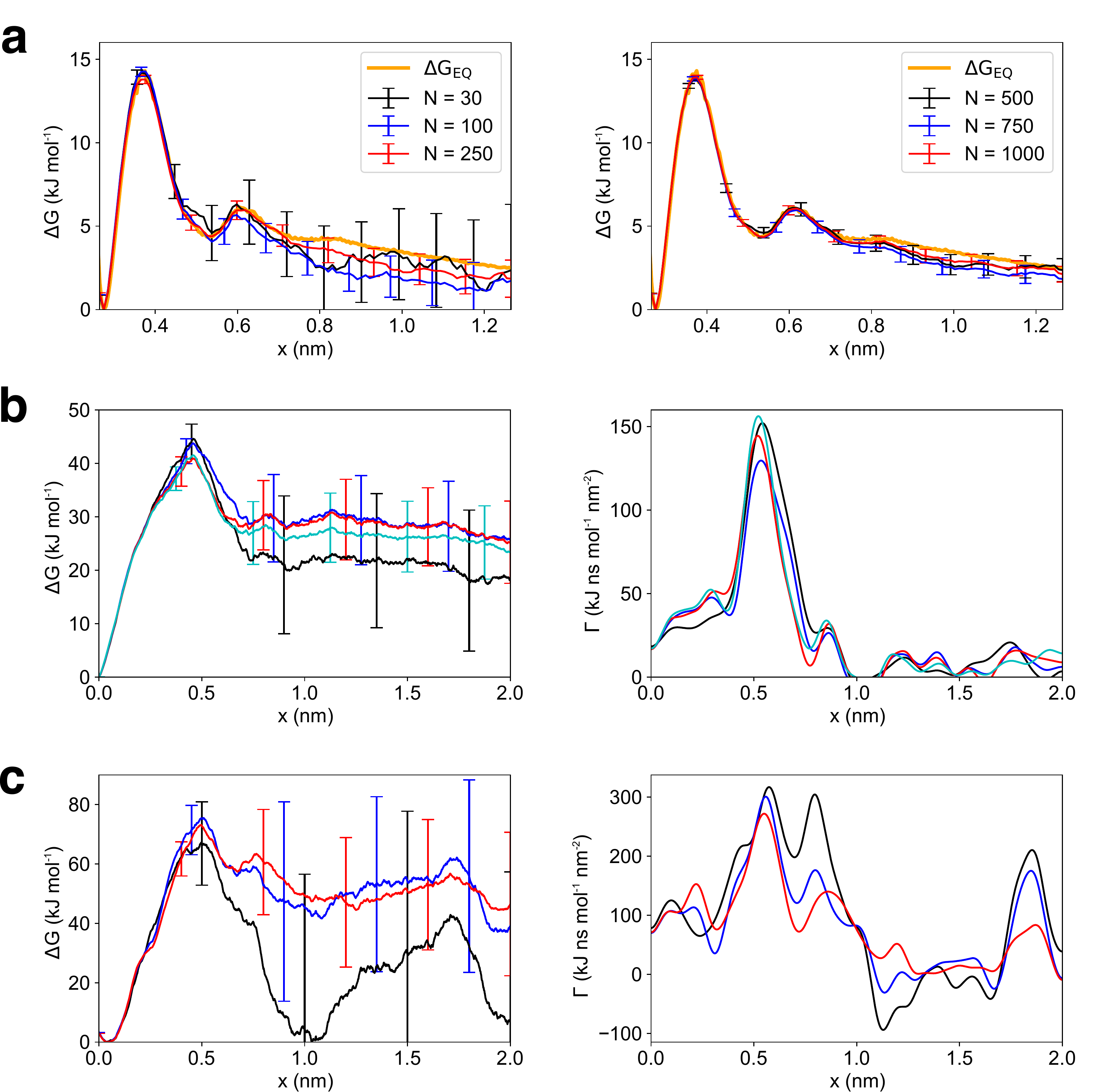} 
\caption{\baselineskip4mm \textbf{Jackknife analysis of free energy
    profiles and Gauss-smoothed friction (\boldmath$\sigma = 40$~ps)},
  as obtained for \textbf{a} NaCl-water, \textbf{b}
  trypsin-benzamidine, and \textbf{c} Hsp90-inhibitor complex. Error
  bars denote the Jackknife standard error obtained for various
  numbers of TMD trajectories (''samples''). Color code in \textbf{b}:
  52 samples in black, 84 in blue, 117 in red, 148 in cyan. Color code
  in \textbf{c}: 30 in black, 50 in blue, 93 in red.}
\label{fig:Jackknife}
\end{figure*}

\newpage
%
%
In a similar vein, we analyzed the convergence of free energies and
friction profiles in our three molecular simulation systems.  To
estimate an error of the free energy profiles, we resort to a
Jackknife (''leave-one-out'') analysis\cite{Efron81}, which represents
a superior method if we do not know the distribution that the free
energy estimate follows. Because the estimate of the free energy
depends on the estimate of both mean and variance of the
nonequilibrium work (see the Theory section in the main text), the
error of the free energy will decrease with the error of the 2\up{nd}
moment of the nonequilibrium work, which is known to converge
slowly\cite{Pearson31}. Consequentially, while we expect the estimate
of the mean free energy to converge comparatively fast, the error will converge
slower.

Figure~\ref{fig:Jackknife} displays the convergence of free energies
and friction profiles of NaCl-water and the two investigated
protein-ligand systems.  In the case of the NaCl system, the free
energy is within 1~$\kT$ of the free energy from unbiased simulations
directly for N=30 simulations, while the error decreases to below
1~$\kT$ within N=100 simulations.  Please see the Supplementary
Information of Ref.~\onlinecite{Wolf18} for the convergence of
NaCl-water friction factors. For the trypsin-benzamidine system, free
energies appear to converge with $\sim$100 trajectories, despite an
error of $\sim$4 $\kT$. In a similar vein, the free energy estimates
for the Hsp90-inhibitor complex change for $\sim$2 $\kT$ after
$\sim$100 trajectories.  A similar convergence appears to apply to friction fields. We therefore expect about 100 trajectories
per path to be sufficient to obtain a sufficiently converged free
energy surface and friction profile for rate prediction.

Furthermore, the binding free energies of $\sim27$~kJ/mol (Trypsin)
and $\sim45$~kJ/mol (Hsp90), calculated as the difference between
$\Delta G$ at $x = 0$~nm and $x=2$~nm, compare well to the standard free
energies of binding $\Delta G^0 = 28.0$~kJ/mol (Ref.~\onlinecite{Guillain70}) and
$40.7 \pm 0.2$~kJ/mol (Ref.~\onlinecite{Amaral17}), respectively,
based on the experimentally measured $K_D = C_0 \exp \left( -\Delta G^0 / \kT \right)$ with the standard reference concentration $C_0 = 1$~mol/l. We are aware that a
distance-based free energy difference as used here and a standard free
energy of binding can differ by several $\kT$ (see
Ref.~\onlinecite{Hall20}). We therefore base the goodness of our
predictions in the main text on protein-ligand complex rates and the resulting
$K_{\rm D}$, respectively.

Lastly, we note that the convergence is best and errors are smallest
directly around the main barrier. Therefore, predictions of unbinding
rates will be more accurate than predictions of binding rates, despite
being the slower and thus actually harder rate to predict.

%
%
\subsection*{Pathway separation}

In the case of trypsin-benzamidine, pathway separation
was performed by employing nonequilibrium principal 
component analysis\cite{Post19} (PCA) 
using a covariance matrix based on protein-ligand contact distances\cite{Ernst15}.
To this end, the PCA included all minimal amino acid-ligand distances that are
found below a cut-off distance of 4.5~\AA\ in any snapshot of all
pulling trajectories. Trajectories were projected onto the first two
principal components and sorted according to pathways by visual
inspection as displayed in Supplementary
Fig.~\ref{fig:pathways}. dcTMD calculations of free energy and
friction were then carried out separately for such bundles of
trajectories. Performing 200 pulling simulations of
trypsin-benzamidine, we found 84 trajectories to constitute the major
unbinding pathway (''middle'' pathway in Fig.~\ref{fig:pathways}), for which free energy and friction profiles were
converged, and whose free energy difference between bound and unbound
state qualitatively agree with the standard free energy of binding
known from experiment (see Supplementary Fig.~\ref{fig:Jackknife}).

For the Hsp90-inhibitor complex, we employed a path separation based on geometric 
distances between individual trajectories\cite{Bray18}. After aligning trajectories with 
the protein's C$_{\alpha}$ atoms as fit reference, we calculated the matrix of means over time of the 
root mean square distance of ligand heavy atoms between individual trajectories. We then applied the 
NeighborNet algorithm\cite{Bryant04} to the matrix to cluster trajectories according to distances. 
From the considered 400 trajectories, the cluster that gave a free energy difference between $x = 0$~nm and $x=2$~nm 
that was closest to the experimental $\Delta G^0$ calculated from the respecitve $K_{\rm D}$\cite{Amaral17} was taken by 93 single trajectories.

\begin{figure*}[ht!]
\centering
\includegraphics[width=0.95\textwidth]{\dirfig/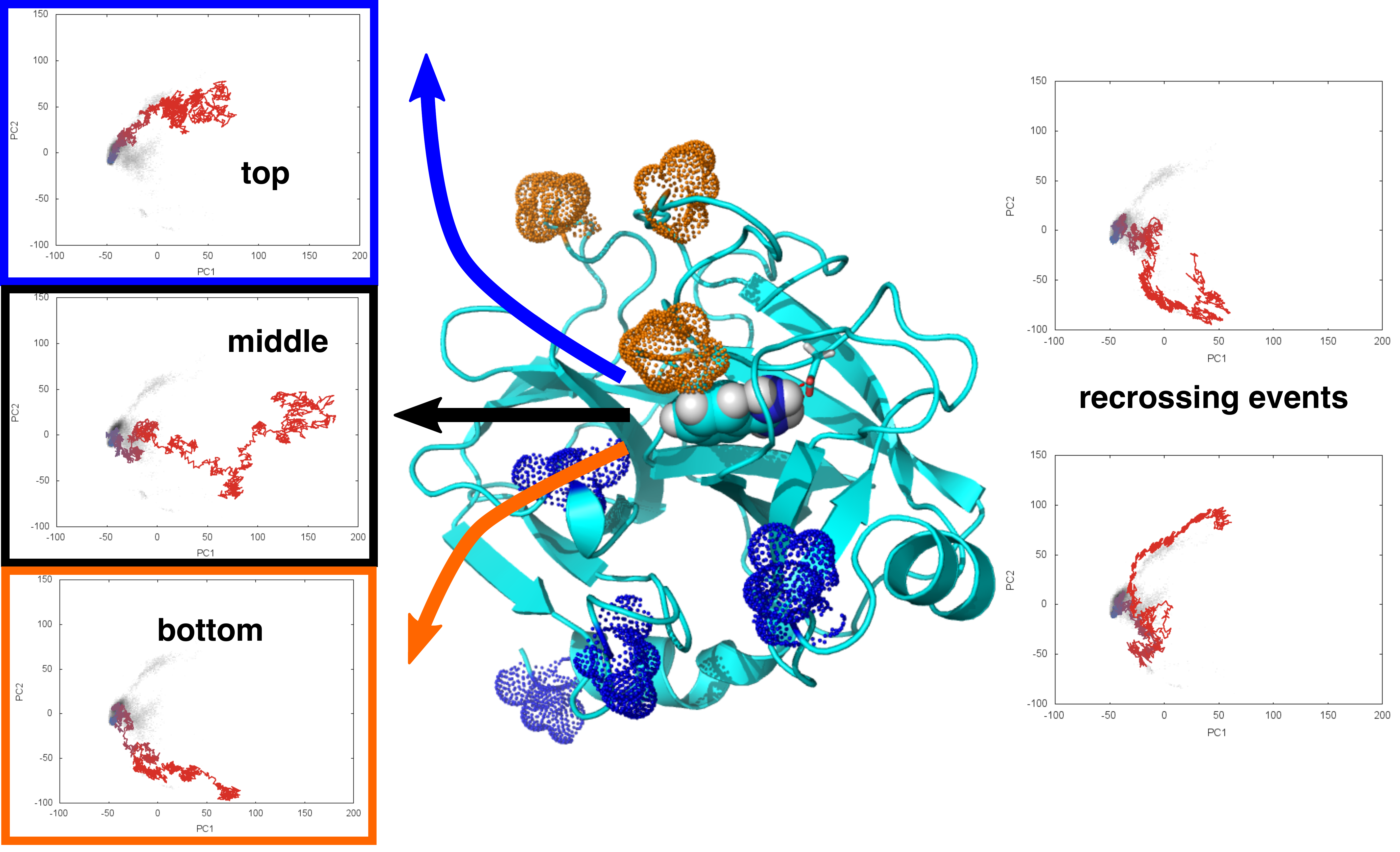} 
\caption{\baselineskip4mm \textbf{Pathways predicted by dcTMD for the
    unbinding of trypsin-benzamidine,} using a pulling velocity of
  $v_c =$ 1 m/s. (Middle) Structure of trypsin (PDB ID
  3PTB)\cite{Marquart83} as cartoon, benzamidine as van der Waals
  spheres, Asp189 in sticks, residues characterizing PC2 as blue and
  orange dots, respectively. Arrows indicate the direction of
  pathways. (Left) Nonequilibrium PCA results display minima in
  nonequilibrium energies as grey shades, corresponding to temporary
  binding sites during unbinding. Projected single representative
  trajectories as lines are colored from blue to red according to time
  evolution. All trajectories start from a central minimum at
  (-50,0). Three main pathways (top, middle, bottom) can be observed:
  the ''top'' pathway passes through the elongated minimum at (0,50),
  the ''middle'' pathway through the central minimum around (-25,0),
  and the ''bottom'' pathway through a shallow minimum at
  ca.~(-10,-70). The ''middle'' pathway is majorly populated. (Right)
  Additionally, trajectories can ''recross'' between the pathways: in
  the presented examples, trajectories first follow the ''middle''
  pathway before crossing over to the ''bottom'' pathway at PC1
  $\approx$ -20~nm (top) or jump between all three pathways at once
  (bottom).}
\label{fig:pathways}
\end{figure*}

\newpage
%
%
\begin{figure*}[ht!]
\centering
\includegraphics[width=0.95\textwidth]{\dirfig/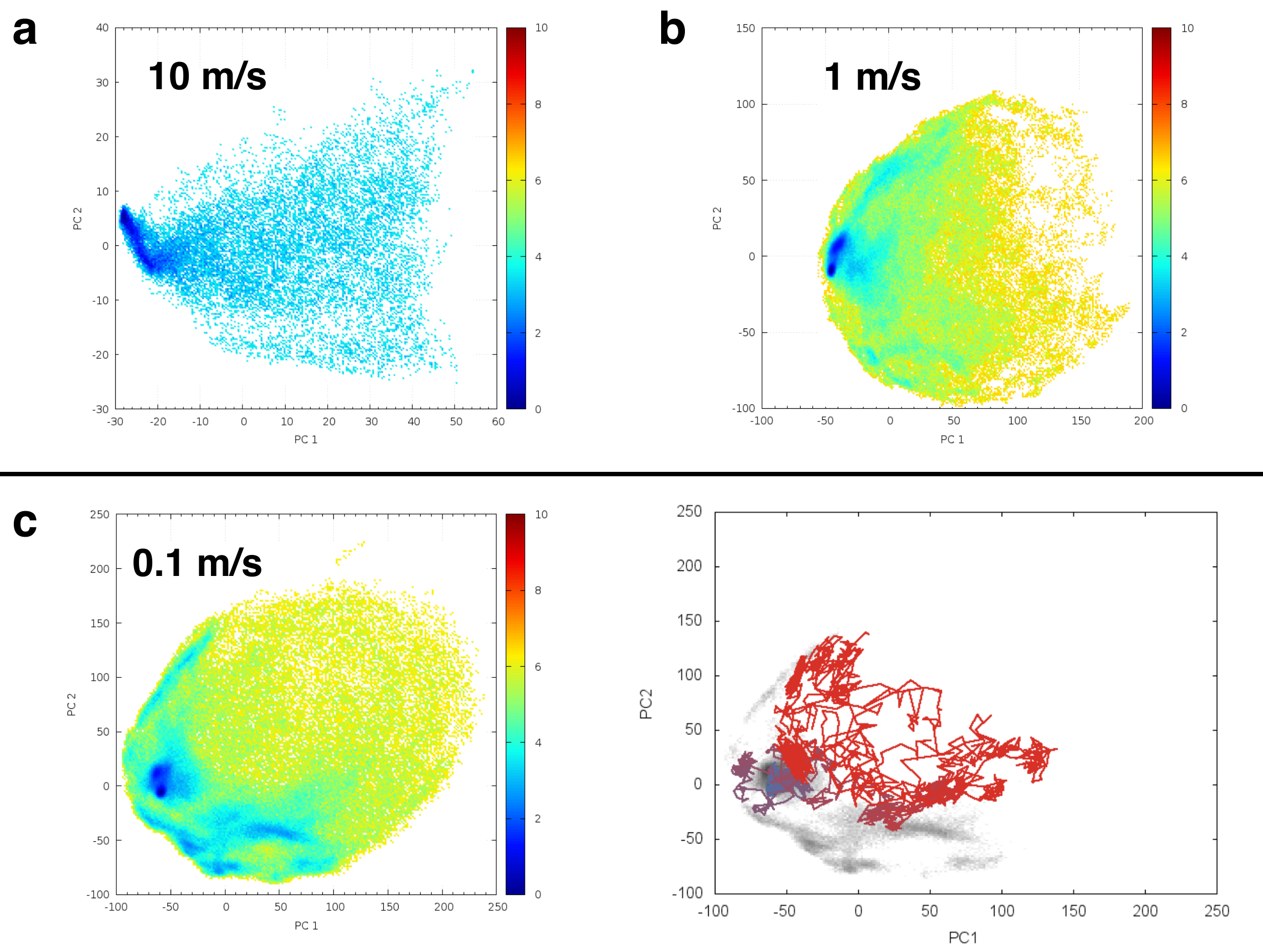} 
\caption{\baselineskip4mm \textbf{Velocity dependence of pathways in
    dcTMD simulations} Nonequilibrium energy landscapes
  $\Delta \mathcal{G}(PC1,PC2) = - k_{\rm B}T P(PC1,PC2)$ obtained
  from dcTMD simulations of the trypsin-benzamidine complex, where
  $P(PC1,PC2)$ is the probability distribution of all dcTMD simulations
  projected on the first two principal components of a contact
  distance PCA \cite{Post19}. For $v_c = 10$ m/s (\textbf{a}), no
  pathways are present, while for 1~m/s (\textbf{b}) structures
  corresponding to pathways appear. At 0.1~m/s (\textbf{c}), a number
  of local energy minima appear, that reflect a multitude of pathways
  present. A pathway separation becomes impossible, as recrossing
  events dominate (see Supplementary Fig.~\ref{fig:pathways}.}
\label{fig:Tryp_PC12}
\end{figure*}

\subsection*{Optimal pulling velocities}

While the results presented above display the evaluation of single
pathways after a successful pathway separation, being able to do so
depends on finding a suitable pulling velocity $v_c$ that allows to
distinguish pathways. To obtain a nonequilibrium work profile that
contains the least dissipative work and at the same time uses the most
likely pathway between states, it intuitively appears to be the best
choice to pull as slowly as possible. Supplementary
Figure~\ref{fig:Tryp_PC12} displays nonequilibrium energy
landscapes\cite{Post19} obtained from dcTMD simulations of the
trypsin-benzamidine complex, using 100 trajectories at $v_c = 0.1$, 1
and 10 m/s. We recognise a tradeoff between the pulling velocity and
the structural resolution of the associated energy landscape. For 10
m/s, we hardly observe any structure in the first two PCs.  Though
this velocity is suited for a scoring of ligands according to
unbinding kinetics\cite{Wolf19c}, obviously a pathway separation
cannot be performed. For 1 m/s, we observe several intermediate states
in the nonequilibrium energy profile along the first two PCs. While
the structure in the first two PCs becomes better resolved in
simulations with 0.1 m/s pulling velocity, sorting trajectories into
unique pathways becomes unfeasible due the dominance of recrossing
events between the pathways (Supplementary Fig.~\ref{fig:pathways}). This finding can
be explained by assuming that single transitions over barriers, as we
try to enforce by applying constraints, take place on a natural timescale 
between pico- and nanoseconds\cite{Dror11a}. Forcing the system
over such barriers on a time scale that is too long, i.e., with a too
slow velocity, causes an artificial stationarity, resulting in the
ligand accessing side states on top of or close to the barrier, which
in equilibrium would not be accessed, and influence from protein
conformational fluctuations. As a consequence, application of dcTMD to
''real world'' systems comes at the price of needing to identify such
''goldilocks'' velocities, with 1 m/s being a good rule of thumb
velocity.

\newpage
%
%
\subsection*{Langevin simulations}

Langevin simulations used the integration scheme by Bussi and
Parrinello\cite{Bussi07a}. Simulations were run for 1~$\mu$s of
simulation time for NaCl at temperatures between 293 to 420~K,
and 10~ms for the two protein-ligand systems, using
temperatures 380 to 900~K for trypsin-benzamidine and 700 to
1350~K for the Hsp90-inhibitor complex. As system mass
$m$, the reduced mass of the NaCl dimer (13.88~g/mol), the
trypsin-benzamidin (120.15~g/mol) and Hsp90-inhibitor (288.73~g/mol)
complexes were used.
System coordinates were written out each 1 ps for NaCl and
each 1 ns for the protein-ligand systems.
\vspace*{-2mm}

\renewcommand{\arraystretch}{0.7}
\begin{table*}[h!]
   \centering
   \caption{\baselineskip4mm Convergence of the Bussi-Parrinello
     integration scheme\cite{Bussi07a} of the Langevin equation with
     respect to the integration time step $\Delta t$ and the 
     mass $m$ used in the Langevin equation. The first table shows
     dissociation and association times obtained from 10 simulations
     of NaCl at $T = 293$~K over 1 $\upmu$s simulated time, as well as
     results from a 1~$\upmu$s long MD simulation. The other two tables
     show results obtained for 5 simulations of trypsin-benzamidine at
     $T = 800$~K over 0.2 ms simulated time, and 5
     simulations of Hsp90-inhibitor complex at $T = 1200$~K over 0.2
     ms simulated time, respectively.
     Errors denote the standard error of the mean.}

   \begin{tabular}{c|cc|cc} 
   \hline
         {\bf ~~~NaCl~~~}   & \multicolumn{2}{c|}{normal mass} &
                                                 \multicolumn{2}{c}{mass $\times$10} \\
      $\Delta t$ (fs) &  $\tau_{\rm diss}$ (ps) & $\tau_{\rm assoc}$ (ps) & $\tau_{\rm diss}$ (ps) & $\tau_{\rm assoc}$ (ps) \\
      \hline
            0.1	&	406 $\pm$ 8 & 	3,248 $\pm$ 66 & 	464 $\pm$ 7 & 	3,580 $\pm$ 50	\\
            0.2	&	404 $\pm$ 8 & 	3,174 $\pm$ 63 & 	467 $\pm$ 12 & 3,528 $\pm$ 80 \\
      0.5		&	411 $\pm$ 8 &	3,166 $\pm$ 65 & 	457 $\pm$ 9 & 	 3,392 $\pm$ 48 \\
      1.0         	&	428 $\pm$ 9 & 	3,157 $\pm$ 65 & 	448 $\pm$ 8 & 	 3,266 $\pm$  78 \\
      2.0        	&      403 $\pm$ 8 & 	3,162 $\pm$ 63 & 	449 $\pm$ 8 &  3,353 $\pm$ 39 \\
      5.0		&	420 $\pm$ 9 & 	3,035 $\pm$ 61 & 	468 $\pm$ 13 & 3,454 $\pm$ 53 \\
      10.0 		&	345 $\pm$ 6 &  2,631 $\pm$ 47 & 	452 $\pm$ 10 & 	3,512 $\pm$ 71 \\
       20.0 		&	249 $\pm$ 5 &  1,907 $\pm$ 25 & 	468 $\pm$ 10 & 	3,475 $\pm$ 67\\
      \hline
      MD	&	124 $\pm$ 6 &	848 $\pm$ 47 	&	124 $\pm$ 6 &	848 $\pm$ 47\\
      \hline 
      \hline
     {\bf Trypsin}       & \multicolumn{2}{c|}{normal mass} & \multicolumn{2}{c}{mass $\times$10} \\
      $\Delta t$ (fs) &  $\tau_{\rm diss}$ (ns) & $\tau_{\rm assoc}$ (ns) & $\tau_{\rm diss}$ (ns) & $\tau_{\rm assoc}$ (ns) \\
      \hline
      0.5		&	111 $\pm$ 2 &	32 $\pm$ 1 & 112 $\pm$ 3 &	33 $\pm$ 1 \\
      1.0         	&	106 $\pm$ 2 & 	32 $\pm$ 1 & 111 $\pm$ 1 &	34 $\pm$  1 \\
      2.0        	&      96 $\pm$ 3 & 	28 $\pm$ 1 & 111 $\pm$ 3 &	33 $\pm$ 1 \\
      5.0		&	61 $\pm$ 2 & 	18 $\pm$ 1 & 112 $\pm$ 5 &	33  $\pm$ 1 \\
      10.0 		&	37 $\pm$ 1 &   10 $\pm$ 1 & 105 $\pm$ 5 &	32 $\pm$ 1 \\
       20.0 		&	 --		 &   --  		 & 95 $\pm$ 4   &	28 $\pm$ 1 \\
      \hline 
      \hline
      ~~{\bf Hsp90}~~      & \multicolumn{2}{c|}{normal mass} & \multicolumn{2}{c}{mass $\times$10} \\
      $\Delta t$ (fs) &  $\tau_{\rm diss}$ (ns) & $\tau_{\rm assoc}$ (ns) & $\tau_{\rm diss}$ (ns) & $\tau_{\rm assoc}$ (ns) \\
      \hline
      0.5		&	628 $\pm$ 25 &	60 $\pm$ 6 & 641 $\pm$ 36 &	61 $\pm$ 5 \\
      1.0         	&	635 $\pm$ 50 & 	60 $\pm$ 4 & 638 $\pm$ 38 &	63 $\pm$ 6  \\
      2.0        	&      567 $\pm$ 18 & 	56 $\pm$ 3 & 636 $\pm$ 35 &	63 $\pm$ 4 \\
      5.0		&	356 $\pm$ 12 & 	36 $\pm$ 2 & 621 $\pm$ 34 &	63 $\pm$ 6 \\
      10.0 		&	212 $\pm$ 8  &   	20 $\pm$ 1 & 608 $\pm$ 21 &	60 $\pm$ 5 \\
      20.0 		&	--   			&   --  		  & 545 $\pm$ 41 &	55 $\pm$ 4 \\
      \hline
   \end{tabular}
   \label{tab:LE_step_NaCl}
\end{table*}

\newpage
For each system, we studied the convergence of the Bussi-Parrinello
integrator with respect to the time step $\Delta t$, see
Table~\ref{tab:LE_step_NaCl}.  We find that NaCl requires an
integration time step of $\Delta t \lesssim 5\,$fs, while the
protein-ligand systems require a significantly shorter time step of
$\Delta t \lesssim 1\,$fs. Owing to the Fourier relation
$\Delta E \Delta t \sim \hbar$, this finding is a consequence of the
larger barrier height $\Delta E$ of the free energy landscape of the
protein-ligand systems.

To test if the dynamics is overdamped, Langevin simulations were
repeated using a mass that is ten times larger than the
normal reduced mass. (Overdamped dynamics neglects the inertia term
and therefore does not depend on the mass \cite{Berendsen07}.) Since
the resulting (un)binding times do not change for the protein-ligand
systems, these systems are clearly overdamped. By using the enhanced
mass, the protein-ligand systems can therefore be integrated by using
a time step of $\Delta t \lesssim 10\,$fs, and therefore require an
order of magnitude less simulation time.
Moreover, the overdamped limit allows us to circumvent the definition
of the effective mass associated with a given pulling coordinate,
which depends on various technical issues such as the definition of
the pulling centers.
On the other hand, NaCl shows
a 15 \% increase of the dissociation and association times, and may be
therefore classified as almost overdamped.

%
%

The gradient of the potential of mean force was approximated as
\begin{align}
\frac{\diff{G}(x)}{\diff{x}} \approx \frac{[ \Delta{G}(x + \Delta x) -
  \Delta{G}(x) ] + [ \Delta{G}(x) - \Delta{G}(x - \Delta x) ]}{2
  \Delta x} 
\end{align}
Input free energy and fiction fields obtained from dcTMD were
smoothed with a Gauss filter ($\sigma = 10$ for NaCl and 40 ps
for protein-ligand systems, respectively). 
%
In some cases, though, friction fields still exhibited
negative values after smoothing, which we found to be a consequence of
not completely converged friction profiles. The problem can be
circumvented by improved sampling or an increased $\sigma = 100$~ps,
which we used in the preparation of Figures 3 and 4 in the main text. As workaround,
we found that using the absolute values $|\Gamma (x)|$ after smoothing
with $\sigma = 40$~ps as input for simulations is sufficient as well,
provided that we have at least 80--100 trajectories available for a
pathway of interest. This workaround was used to prepare fields for
protein-ligand Langevin simulations. For $x$, we used a resolution of
1~pm.  For compensation of data borders, we employed ''fully
reflective'' boundary conditions: If the system jumped over a boundary
$x_{\rm max}$ at any time step by a distance $a$, it was put back to
$x = x_{\rm max}-a$, and its velocity sign reversed.

Mean waiting times $\tau$ were calculated by
defining geometric cores\cite{Nagel19}: For NaCl, the free energy
surface was separated into the bound state $x < 0.31$ nm and unbound
state $x > 0.43$. For trypsin-benzamidine, we used a bound state
$x < 0.3$~nm and unbound state $x > 0.6$~nm, while for the
Hsp90-inhibitor complex, we applied cores of $x < 0.3$~nm and
$x > 0.9$~nm. As the native units of $k_{\rm on}$ are
$s^{-1} \, M^{-1}$, all according binding rates were scaled by a pulling coordinate-dependent
reference concentration $C = 1 / \left( \frac{4}{3} \pi x_{\rm ref}^3 \right)$ for one ion or protein-ligand pair
with $x_{\rm ref} = x_{\rm end} - x_0$, amounting to a molarity of
0.2 M for NaCl Langevin simulations and 50 mM for protein-ligand
Langevin simulations.

\begin{figure*}[tb!]
\centering
\includegraphics[width=0.45\textwidth]{\dirfig/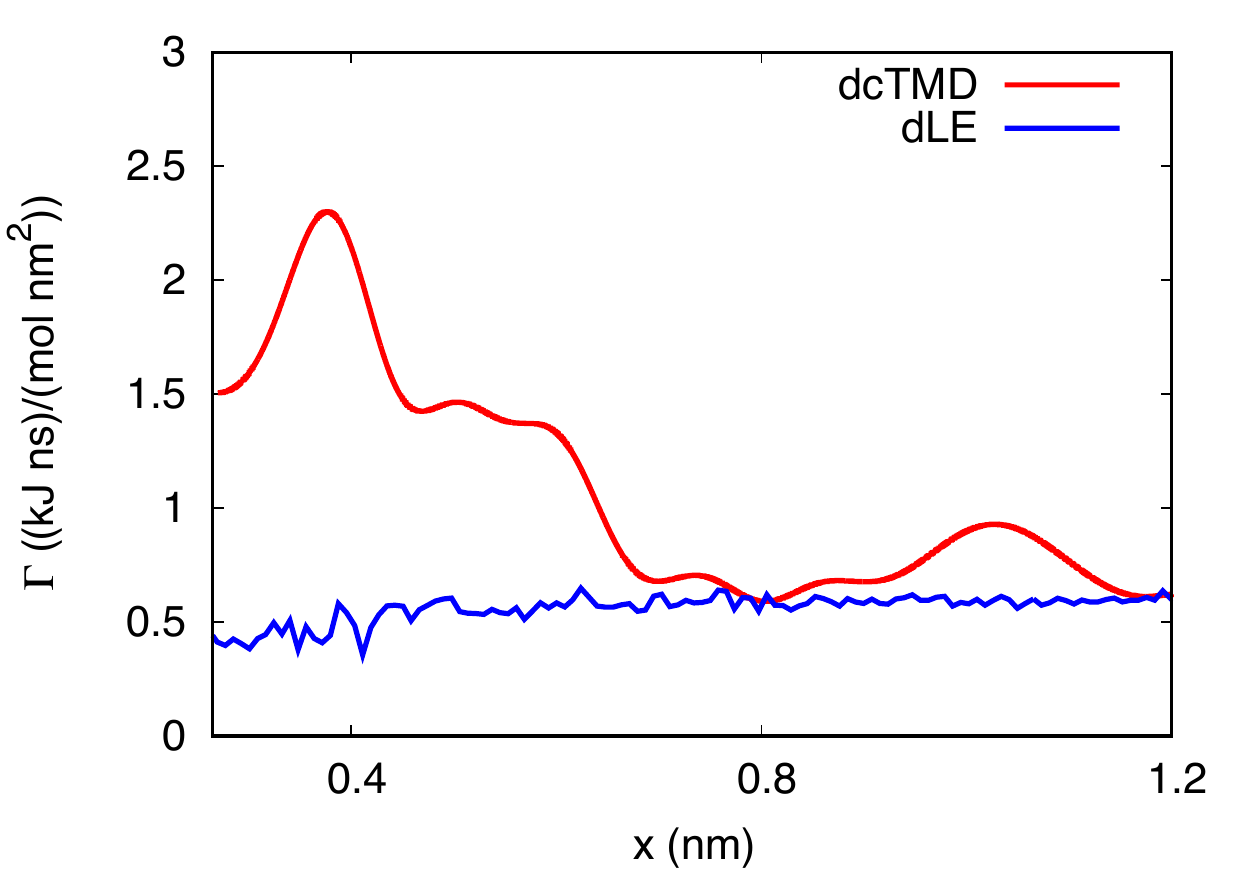} 
\caption{\baselineskip4mm \textbf{Comparison of friction estimates.}
  Friction calculated from a data-driven Langevin equation
  \cite{Schaudinnus16} in comparison to results from
  dcTMD, obtained for the example of NaCl in water.}
\label{fig:dLE}
\end{figure*}

To demonstrate the effect of constraints used in dcTMD, Supplementary
Figure~\ref{fig:dLE} displays a comparison of friction fields of the
NaCl-water system obtained from dcTMD and a data-driven Langevin
equation (dLE)\cite{Schaudinnus16}. The latter uses no constraints,
but calculates the friction via a local average
\cite{Schaudinnus16}. The dLE was applied to 200 ns unbiased
equilibrium MD data and the consistency of the $\Gamma$ estimate was
verified by comparing MD and dLE dynamics. While both fields differ in
the details of shape, the dLE field reaches approximately the final
friction value from dcTMD in the unbound state at maximal $x$.
On average, we find that dcTMD clearly overestimates the friction,
which presumably is caused by the moving position constraints
\cite{Daldrop17}. 

%
%
\subsection*{Uncertainty for prediction of rates with temperature
  boosting}

To estimate the extrapolation error of $T$-boosting, we assume that
the waiting time $\tau$ of an unbinding or binding process is
exponentially distributed,
\begin{equation}
P(\tau)=\frac{1}{\langle \tau\rangle}e^{-\frac{\tau}{\langle \tau\rangle}},
\end{equation}
where $\mean{\tau}$ is a function of the temperature $T$. Hence the expectation
$\bar{\tau}$ is given by its mean $\langle \tau\rangle$ plus/minus the
error of the mean 
\begin{equation}
\bar{\tau}(T)=\langle \tau(T)\rangle\pm\frac{\langle
  \tau(T)\rangle}{\sqrt{N(T)}}, 
\end{equation}
where $N(T)$ denotes the number of simulated transitions.
By changing to the dimensionless rate $k=t_0/\langle \tau\rangle$
(with $t_0$ being some timescale, e.g., ns) and employing Gaussian
error propagation to lowest order\cite{Hughes10}, we obtain
accordingly  
\begin{equation}
\ln{(\bar{k}(T))}=\ln{(k(T))}\pm \frac{1}{\sqrt{N(T)}} .
\end{equation}
Due to the rate expression
$k\!\propto\! e^{-\Delta G/k{\rm _B}T}$
with transition state energy $\Delta G$, $\ln(k)$ depends linearly on
$1/T$, i.e., 
\begin{equation}
\ln(k(T))=\frac{a}{T}+b .
\end{equation}
Linear regression theory\cite{Hughes10} yields estimates for $a$ and
$b$ as well as uncertainties 
\begin{equation}
\sigma_{b}=\sqrt{\frac{\sum_i\frac{N(T_i)}{T_i^2}}{\Delta}}
\end{equation}
and
\begin{equation}
\sigma_{a}=\sqrt{\frac{\sum_iN(T_i)}{\Delta}}
\end{equation}
with
\begin{equation}
  \Delta=\sum_iN(T_i)\sum_i\frac{N(T_i)}{T_i^2}-
  \left(\sum_i\frac{N(T_i)}{T_i}\right)^2  ,
\end{equation}
where ${T_i}$ denotes a discrete set of temperatures at which
simulations are performed. Employing error propagation, we estimate
the uncertainty of $\ln(k)$ at $T_{\text{ref}}=300$ K as 
\begin{equation}
  \sigma_{\ln(k(T_{\text{ref}}))}=\sqrt{\left(\frac{\sigma_a}
      {T_{\text{ref}}}\right)^2+\sigma_b^2}.
\end{equation}
This yields for the desired relative uncertainty of the average waiting time
\begin{equation}
\bar{\tau}(T_{\text{ref}})=\langle\tau(T_{\text{ref}})
\rangle\pm\langle\tau(T_{\text{ref}})
\rangle\cdot\sigma_{\ln({k}(T_{\text{ref}}))}  .
\end{equation}

To illustrate the typical magnitude of this uncertainty, we assume
that we perform $10$ Langevin simulations of length $t_{\text{LE}}$ at
different temperatures $T_i$ ($i=0,$ ... $,9$)
\begin{equation}
T_i=T_0+i\left(25\frac{T_0}{T_{\text{ref}}}\right)\text{K}.
\end{equation}
The first three temperatures ($T_0$, $T_1$ and $T_2$) are chosen such
that we observe $\approx10^2$ transitions during the simulation time
$t_{\text{LE}}$. Similarly, we assume to observe $10^3$ transitions at
$T_3$, $T_4$ and $T_5$, $10^4$ transitions at $T_6$, $T_7$ and $T_8$
and $10^5$ transitions at $T_9$.
Using the case of $T_0=T_{\text{ref}}=300$ K, since we assumed
$10^2$ transitions at $T_0=T_{\text{ref}}$ during the Langevin
simulation time $t_{\text{LE}}$, the observed rate at $300$ K is
$k=10^2/t_{\text{LE}}$. Choosing $t_{\text{LE}}=5$ ms, the
corresponding rate is $k(300\text{K})={1}/{50\mu\text{s}}$, and we obtain for the error of
the rate $\sigma_{\ln({k}(T_{\text{ref}}))}=7.7\%$.
Alternatively, when we assume that we need to choose $T_0\approx450$ K in order to
achieve $10^2$ transitions, employing the boosting relation (4) in the
main text and $t_{\text{LE}}=5$ ms, the observed rate at $300$ K is
$k(300\text{K})=0.063$ ms$^{-1}$ with an error of $10.6\%$. 

Considering our Langevin simulations of trypsin described in the main text, 
where we used $T$-boosting
at $13$ temperatures from 380--900~K, the error at 300~K is estimated to be
$\sigma_{\ln({k}(T_{\text{ref}}))}=3.3\%$. For Hsp90, the system with the highest considered barrier, 
we obtain $\sigma_{\ln({k}(T_{\text{ref}}))}=11.0\%$ at 300~K using Langevin simulations
at 14 temperatures from 700--1350~K.
As the overestimation of friction factors due to usage of constraints
(see main text) results in a underestimation of rates by a factor of
$\sim 4$, and as the error of free energy profiles enters Eq. $[4]$ in
the exponent, the extrapolation error due to $T$-boosting can easily
be made negligible in all practical cases.

\end{document}